\documentclass[
 aip, apl, amsmath, amssymb, reprint]{revtex4-1}
\usepackage{graphicx}
\usepackage{dcolumn}
\usepackage{amsmath}
\usepackage{bm}
\usepackage{subcaption}
\usepackage[justification=raggedright]{caption}
\usepackage{float}
\usepackage[demo,abs]{overpic}

\begin{document}

\preprint{AIP/123-QED}

\title{Nonlinear effects in high-intensity focused ultrasound power transfer systems}

\author{Aarushi Bhargava}

\author{Vamsi C. Meesala}%
\affiliation{Department of Biomedical Engineering and Mechanics, Virginia Tech, Blacksburg, VA, $24061$, USA
}

\author{Muhammad R. Hajj}
\affiliation{Department of Civil, Environmental and Ocean Engineering, Davidson Laboratory,  Stevens Institute of Technology, Hoboken, NJ, $07030$, USA}

\author{Shima Shahab}
\email{sshahab@vt.edu}
\affiliation{Department of Mechanical Engineering, Virginia Tech, Blacksburg, VA, $24061$, USA}
\altaffiliation[Also ]{Department of Biomedical Engineering and Mechanics, Virginia Tech, Blacksburg, VA, $24061$, USA}

\begin{abstract}
In the context of wireless acoustic power transfer, high intensity focused ultrasound technology aims at the reduction of spreading losses by concentrating the acoustic energy at a specific location. Experiments are performed to determine the impact of nonlinear wave propagation on the spatially resonant conditions in a focused ultrasonic power transfer system. An in-depth analysis is performed to explain the experimental observations. The results show that the efficiency of the energy transfer is reduced as nonlinear effects become more prominent. Furthermore, the position of the maximum voltage output position shifts away from the focal point and closer to the transducer as the source strength is increased. The results and analysis are relevant to the development of novel efficient ultrasonic power transfer devices when using focused sources. 
\end{abstract}
\keywords{Wireless power transfer, Acoustic power transfer, HIFU, Nonlinear acoustics, Piezoelectric}

\maketitle

Ultrasonic power transfer (UPT) has emerged as a promising technology to wirelessly power devices or sensors \cite{cochran1988external,shi2016mems} including through-wall power transmission \cite{bao2008high,graham2011investigation,hu2003transmitting} and for wireless data delivery\cite{kawanabe2001power,sanni2012inductive}. A UPT system consists of a piezoelectric transducer that converts the input electrical power to vibration-induced acoustic waves to be received by a piezoelectric disk that, in turn, converts acoustic-induced vibrations to electric power    \cite{shahab2014contactless,ozeri2010ultrasonic,shahab2015ultrasonic,meesala2020modeling,meesala2020acoustic}. Acoustic waves are biologically safe and have short wavelengths that allow for effective integration in small transducers and receivers\cite{roes2012acoustic,awal2016state}.  Because of these advantages, UPT has been favored over more traditional wireless energy-transfer technologies that employ electromagnetic waves such as capacitive, inductive, and microwave-based methods \cite{kim2012wireless,li2012magnetic,kar2015bi}. On the other hand, the spreading losses from planar and spherical acoustic sources is a key challenge that has limited the implementation of UPT technology \cite{awal2016state}. This study proposes to combine high intensity focused ultrasound (HIFU) with UPT. The expectation is that HIFU would increase the efficiency of the HIFU-UPT system as the propagating acoustic energy from the source is concentrated over a small spatially localized spot where a receiver can be placed to receive maximum acoustic power \cite{cui2011enhanced,cui2013laser,bhargava2017focused}.

When considering an HIFU-UPT system, it is important to note that wave distortion is more complex and pronounced in a focused pressure field due to combined nonlinear and diffraction effects, especially in the focal region. Additionally, reflections from the surfaces of the HIFU source and the receiver result in the formation of standing waves \cite{rudenko2001nonlinear,vanhille2005numerical,coppens1968finite,andrade2016acoustic,bassindale2014measurements}  with antinode locations that exhibit maximum localized values of the acoustic pressure.  As the amplitude of the source excitation is increased, these waves become increasingly nonlinear whereby more energy is transferred from the excitation frequency to its higher harmonics. This energy transfer also varies with distance from the source  \cite{hamilton1998nonlinear}. Depending on the geometry, material properties, and location of the receiving piezoelectric disk, it may not be possible to harvest energy from the harmonic components, which reduces the capability of harvesting the voltage from a nonlinear field. Clearly, different phenomena impact the dynamics of spatially resonant focused acoustic field in HIFU-UPT systems and play crucial role in determining the efficiency and maximum voltage output position (MVOP).

\begin{figure}
    \begin{center}
			\includegraphics[width=0.375\textwidth]{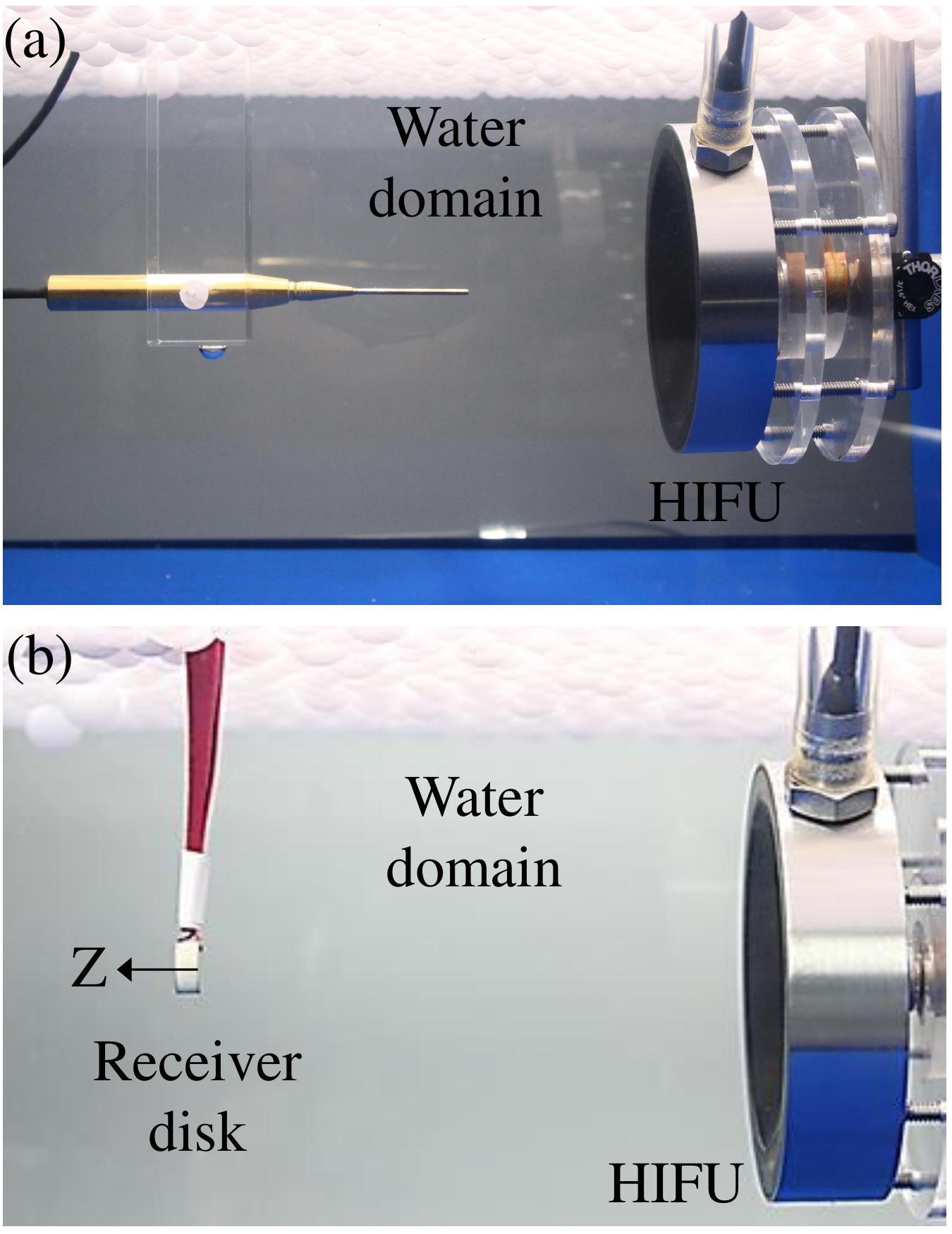}
		\end{center}
		\vspace{-12pt}
	\caption{Pictures of the experimental setup showing suspended (a) hydrophone to map the HIFU pressure field and identify the focal point and (b) piezolectric disk used to harvest voltage from the HIFU source. The positive direction of the $Z$ axis is away from the HIFU source. The focal point is marked as $z=0$.}
	\label{fig1}
	\vspace{-18pt}
\end{figure}

The objectives of the performed experiments are to investigate the dynamics impacting the MVOP under focused and nonlinear spatially resonant acoustic conditions in an HIFU-UPT system and assess its efficiency when operating in linear and nonlinear pressure fields. In the experiments, the HIFU transducer manufactured by Sonic Concepts® that operated at a frequency of $0.5$ MHz was placed at one end of a $61.5 \times 31.8 \times 32.5$ cm$^3$ water tank. The tank was lined with Precision acoustics F$28$ absorbing sheets to prevent reflections from the tank walls. The transducer was actuated using a Keysight $33500$B signal generator and E\&I amplifier. The tank was filled with deionized and degassed water to prevent electrical short-circuiting and cavitation. In the first set of experiments, a Precision Acoustics $1$ mm needle hydrophone was suspended using a positioning system, as shown in as shown in figure \ref{fig1}a, to map the HIFU pressure field and identify the focal point. To record the pressure field, the HIFU was operated at $0.5$ MHz with a burst signal of $80$ $\mu$s and burst periods of $5$ ms. In a second set of experiments, the hydrophone was replaced with a $3.9$ mm thick and $9.5$ mm wide piezoelectric disk manufactured by APC as shown in figure \ref{fig1}b. The HIFU operating conditions were changed to $1.4$ ms of burst signal with $1$ s of burst period to record the spatially resonant acoustic-electroelastic response of the disk. The selected duration was long enough to form a standing wave pattern between the transmitter and receiver. The receiver disk was connected to an optimum load resistance of $1$ kOhm, determined from a different set of experiments that included sweeping over a range of load resistances between $10$ Ohm and $1$ MOhm.  

\begin{figure}
		\begin{center}
			\includegraphics[width=0.48\textwidth]{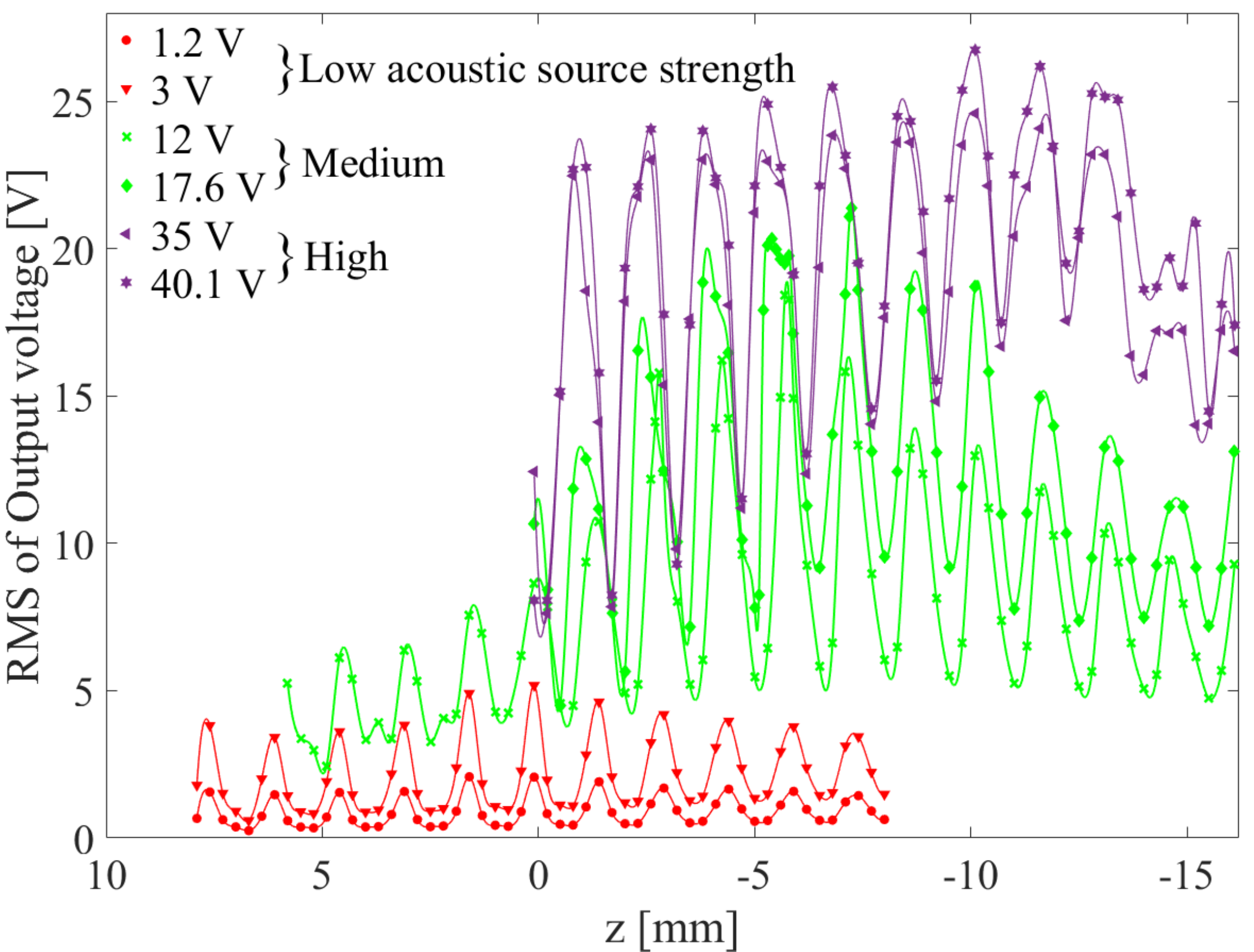}
		\end{center}
		\vspace{-12pt}
		\caption{Output voltage along the z-axis for different levels of input voltage to the HIFU. The different colors denote variations in the pressure field from linear to weakly linear to saturation as the excitation is increased from low to medium to high amplitudes, respectively.}
		\label{fig2}
		\vspace{-18pt}
\end{figure}

Figure \ref{fig2} shows the root mean square (RMS) of the harvested voltage by the disk at different positions along the z-axis for relatively low, medium and high excitation amplitudes. The observed fluctuations are due to the standing wave field with local maxima occurring at $\lambda/2$ separations where $\lambda=c_0/f_0$ is the ratio of the  the speed of sound in the medium, $c_0$, to the excitation frequency, $f_0$. The effects of nonlinear excitation on the pressure field were determined by increasing the excitation level. At relatively low excitation levels, between $1.2$ and $3$ V, the respective maximum output voltages are approximately $2.1$ and $5.2$ V. In both cases, this maximum is located at the focal point, $z=0$, as expected since the maximum energy concentration of a linear pressure field is at the focal point. As the source excitation level is increased to relatively medium levels between $12$ and $17.6$ V, the respective maximum output voltages increase to values between 18.4 and 21.4 V. Moreover, the MVOP shifts towards the HIFU source. The decrease in the ratio of the output to the input voltage and the shift in the MVOP indicate that nonlinear effects are significant at these excitation levels. Increasing the input voltage to higher levels between $35$ and $40.1$ V yields local maximum output values between $24.6$ and $26.7$ V at many locations between the source and the receiver. The further reduction in the ratio of the output to input voltage when compared to the cases of medium excitation levels and the broadening of the MVOP indicate that saturation may be governing the pressure field at these excitation levels. The observed reduction in the ratio of output to input voltages and the shift in the MVOP as the input voltage is increased contradict the expectation that maximum energy of the pressure field can be realized at the focal point at high excitation levels. 

\begin{figure}
    \begin{center}
			\includegraphics[width=0.48\textwidth]{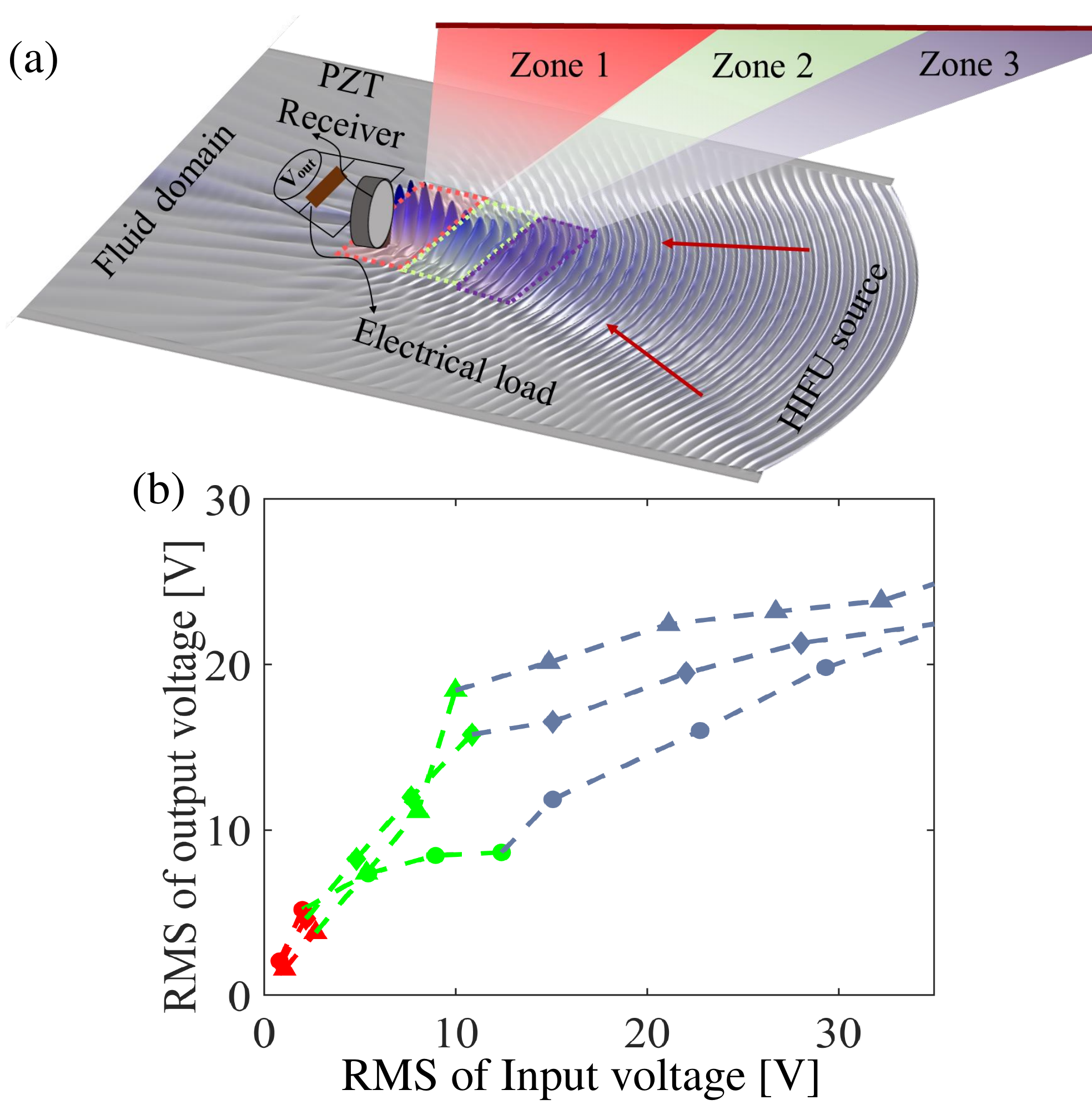}
	\end{center}
	\vspace{-12pt}
	\caption{(a) Schematic of the HIFU actuated piezoelectric receiver in a UPT system. The receiver is located at the focal point of the transducer and is connected to an electrical load resistance. The excitation is along the polarity (thickness) direction, (b) Variations in receiver response for different RMS amplitudes of source excitation in the different zones. The  circular, diamond, and triangular symbols represent the maximum output voltage observed respectively in zones 1, 2 and 3. The colors denote the level of excitation as shown in Fig. \ref{fig2}.}
	\label{fig3}
	\vspace{-18pt}
\end{figure}

To further understand the acoustic influence on voltage response for different levels of the excitation voltage, we provide a schematic outlining three zones of the HIFU-UPT system as shown in figure \ref{fig3}a. The planar pattern shows expected spatial variations of the acoustic pressure as a consequence of the standing waves formed by the constructive and destructive interference of incident and reflected waves between the HIFU source and receiver surfaces. Based on this pattern, we highlight zone $1$  between $0.1$ and $-0.1$ cm,  zone $2$  between $-0.1$ and $-0.7$ cm, and zone $3$  between  $-0.7$ and $-1.45$ cm on the axial axis and compare the RMS of the maximum output voltage by the receiver in these zones in figure \ref{fig3}b. The results show different linear and nonlinear responses in the three zones. At relatively low excitation voltages, up to $3$ V, the highest response, as noted by the red circles, is in zone $1$.  The voltage outputs in zones $2$ and $3$, signified by the red diamond and triangular symbols, are slightly lower. Increasing the voltage to a medium range between $3$ and $12$ V increases the output voltage in all zones. However, the highest response increase occurs in zone $2$ represented by the green diamond, with significantly lower response in zone 1 as shown by the green circles. Increasing the voltages to values between $12$ and $35$ V causes an increase in the output voltage in all zones. However, the highest voltage output is in zone $3$ represented by the purple triangular symbol. It is important to note that the output voltage tend to saturate in all zones as the input voltage is increased to $35$ V indicating that saturation conditions have started to develop for these input values. 

 The distribution of the time-varying acoustic pressure in the HIFU field and in the absence of a receiver is governed by the Khokhlov Zabolotskaya-Kuznetsov (KZK) equation written as \cite{hart1988nonlinear}
\begin{equation}
    \frac{\partial^2p}{\partial z \partial \overline{t}}=\frac{c_0}{2}\nabla_{\perp}^2p+\frac{\delta}{2c_0^3}\frac{\partial^3p}{\partial t^3}+\frac{\beta}{2\rho c_0^3}\frac{\partial^2 p^2}{\partial \overline{t}^2} 
    \label{eq1}
\end{equation}
where $p$ is the acoustic pressure, and $\overline{t}$ is the retarded time defined as $\overline{t}=t-z/c_0$ with $t$ as time. The Laplacian operator is defined as $\nabla_{\perp}^2=\partial^2/\partial r^2 + (1/r)\partial/\partial r$ where $r$ is the radial distance along radial axis, $R$. The first term on the right-hand side of Eq. \ref{eq1} represents the diffraction due to focusing. Changing  geometrical parameters such as aperture radius or radius-of-curvature of the HIFU source modifies the diffraction effects, which consequently alters the dimensions of the focal zone \cite{canney2008acoustic}. The second term represents thermo-viscous medium attenuation determined by the diffusivity of sound in a specific medium $\delta$. The third term represents nonlinear effects with $\beta$ denoting the nonlinearity coefficient. These effects are due to the inherent nonlinearity of the medium, which distorts the waveform. This distortion takes place when the phase speed of the particles in the compression, or high pressure region, of the waveform becomes higher than that of the particles in its rarefaction, or low pressure region. In terms of energy content of the waveform, the distortion is associated with energy transfer from the fundamental frequency to higher harmonics \cite{hamilton1998nonlinear}. Because of nonlinear effects of the medium, the level of the harvested voltage is limited to a saturation value where the pressure at a specified location reaches a maximum that is independent of the input excitation to the source. This condition is referred to as acoustic saturation condition \cite{canney2008acoustic,duck2002nonlinear,muir1980prediction}. Clearly, assessing the nonlinear effects is important to prevent the operation of the UPT system under acoustic saturation that can lead to a decrease in the efficiency of the system. 

To discern the effect of acoustic nonlinearity on the performance of the piezoelectric disk used in the experiments, we perform finite-element simulations combining Eq. \ref{eq1}, and the principles of acoustic-structure interaction physics \cite{fahy2007sound} and electroelastic dynamics \cite{leo2007engineering} based on the model by Bhargava and Shahab\cite{bhargava2020contactless}. In these simulations, the disk used in the experiments and operating under a load resistance of $1$ Ohm was considered. The disk was placed at the focal spot of the HIFU transducer whose operating parameters were identified by experimentally validating its response with Eq. \ref{eq1} \cite{bhargava2020contactless}. This disk was chosen as it possessed a thickness mode near $0.5$ MHz, which is the operating frequency of the HIFU transducer. The simulations were performed according to the specifications in Bhargava and Shahab\cite{bhargava2020contactless} for linear and nonlinear acoustic excitation conditions, in the absence of material nonlinearities of the disk and acoustic spatial resonance. The linear condition is defined by the presence of frequency component only at $0.5$ MHz in the frequency spectrum of the pressure field. Under nonlinear excitation conditions, the spectrum contains the excitation frequency and its harmonics. 

\begin{figure}
    \begin{center}
		\includegraphics[width=0.4\textwidth, height=0.55\textwidth]{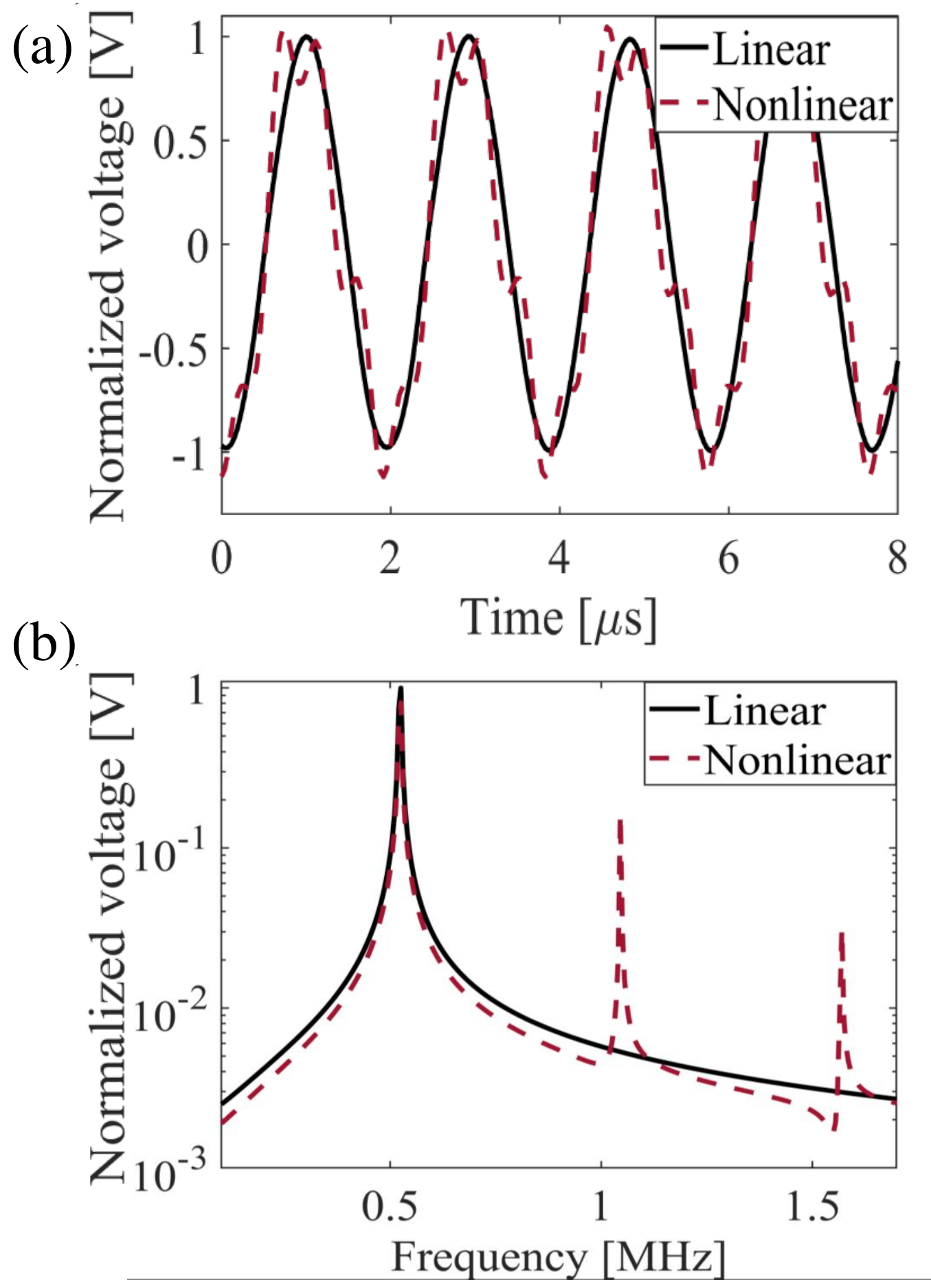}
	\end{center}
	\vspace{-12pt}
	\caption{(a) Time series and (b) spectra of simulated responses of linear and weakly nonlinear (non-saturation conditions) of a piezoelectric receiver using the KZK equation. Plots are normalized with the maximum output voltage obtained from linear acoustic excitation.}
	\label{fig4}
	\vspace{-18pt}
\end{figure}

Plots of the time series and corresponding spectra of the responses of the disk under linear and nonlinear conditions are presented in Fig. \ref{fig4}a and Fig. \ref{fig4}b. The plots are normalized with the maximum output voltage obtained from linear acoustic excitation. The plots show that the disk has the same normalized response amplitude under linear and nonlinear acoustic excitation conditions. A comparison of the frequency spectrum of the two voltage responses shows that the higher frequency components in the voltage response are at least one order of magnitude smaller than that of the fundamental component, which highlights that only the fundamental mode of the disk has a significant contribution to the output voltage. This result is a consequence of higher structural modes of the disk not coinciding with the acoustic harmonics as explained by Bhargava and Shahab\cite{bhargava2020contactless}. However, if a larger number of non-negligible acoustic harmonics and structural mode frequencies coincide, the higher frequency components of the voltage response would be comparable in magnitude to the fundamental component\cite{bhargava2020contactless}. Still, in the current measurements, only the fundamental pressure component is impacting the response of the piezoelectric disk, which explains the drop in the ratio of the output to input voltage as the input voltage is increased. This is because energy transferred from the fundamental to the higher acoustic harmonics is not picked up by the disk.	

\begin{figure}
    \begin{center}
		\includegraphics[width=0.4\textwidth, height=0.55\textwidth]{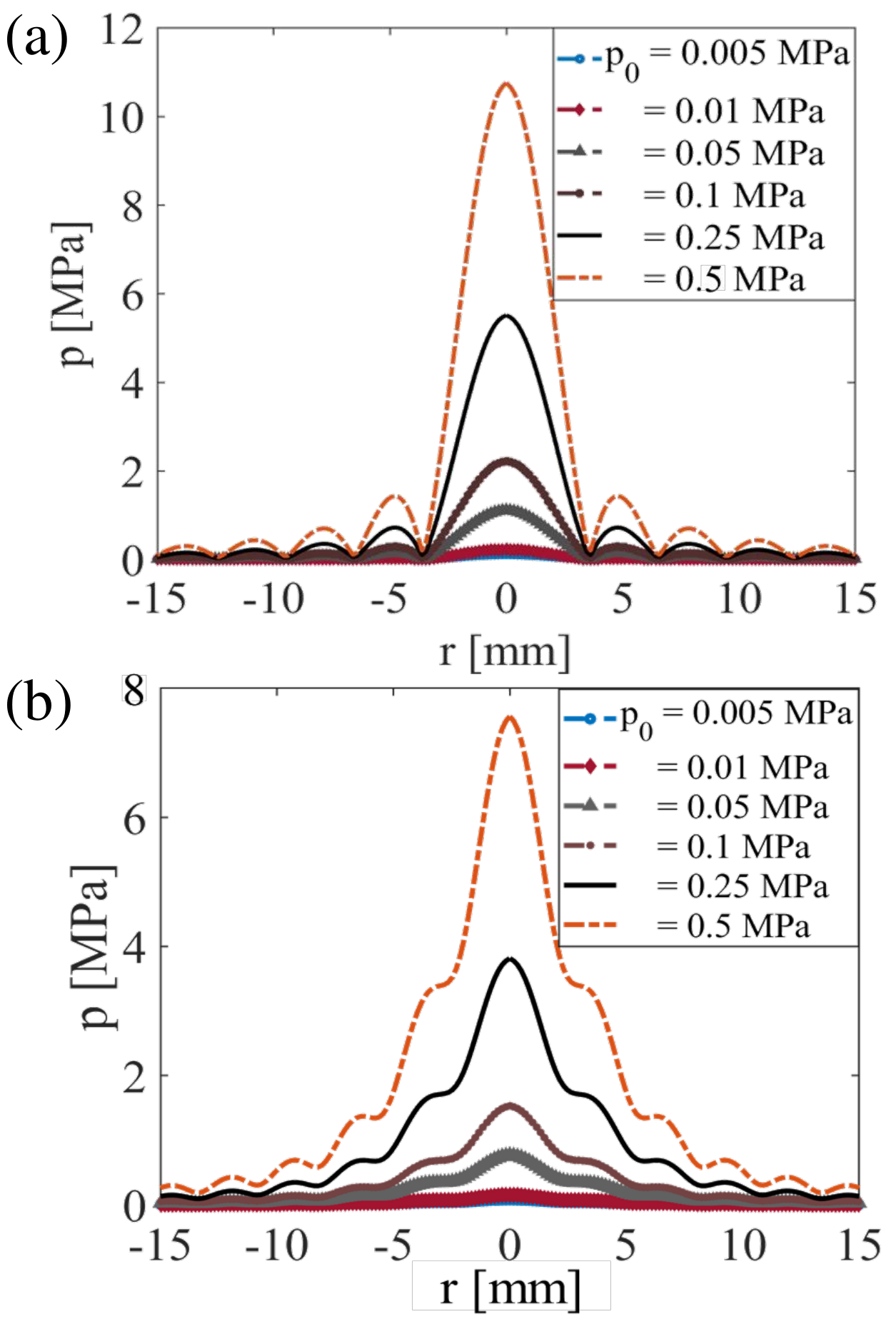}
	\end{center}
	\vspace{-12pt}
	\caption {Variations along the radial axis of the amplitude of the fundamental component of the pressure field at (a) the focal point $z=0$ and (b) $z=-10$ mm.}
	\label{fig5}
	\vspace{-18pt}
\end{figure}

Having determined that most of the acoustic excitation of the receiving disk is due to the fundamental component, the variations of this component along the transverse (radial) axis is analyzed next. The transverse pressure field evaluated using Eq. \ref{eq1} in the absence of disk for excitation levels spanning over two orders of magnitude at  $z=0$ and $z = -10$ mm are respectively presented in figures \ref{fig5}a and \ref{fig5}b.  Because the rate of energy transfer from the excitation frequency to the harmonics increases as the amplitude of the excitation is increased \cite{hamilton1998nonlinear,hart1988nonlinear}, the rate of increase in the amplitude of the fundamental component of the pressure field is less than the rate of the increase in input source level denoted by $p_0$. Furthermore, the accumulation of nonlinear effects with distance of the pressure field leads to a larger percentage reduction in the fundamental amplitude at the focal point relative to the amplitude at $z=-10$ mm along the axis, $r$=0. In addition, the  diffraction effects increase the width of the main lobe at positions away from the focus. These variations cause the effective acoustic force on the disk, which is the surface integral of the amplitude of the fundamental component along the radial axis and within the disk area, defined by a radius of $4.5$ mm, to be larger at $z$=-10 than at the focal point $z$=0. This variation explains the reduction of voltage response of the disk in zone $1$ and the larger response in zone $3$, with increase in excitation amplitude in Fig. \ref{fig3}b. 

The effects of medium nonlinearity and acoustic spatial resonance on the performance of an HIFU-UPT system were experimentally investigated. The results point to a shift in the maximum voltage output position away from the focal point and a reduction in the system's efficiency as the excitation level is increased. In-depth analysis and simulation of the governing equation show that the transfer of energy from the fundamental to higher harmonics leading to acoustic saturation and the radial distribution of the acoustic energy determined by the nonlinear effects play a key role in shifting the MVOP and in reducing the system's efficiency. 
\vspace{-15pt}
\section*{AUTHOR'S CONTRIBUTIONS}
\vspace{-10pt}
A. Bhargava and V. Meesala contributed equally to this work.
\vspace{-15pt}
\section*{ACKNOWLEDGEMENT}
\vspace{-10pt}
This work was supported by the National Science Foundation Grant No. ECCS-$1711139$.

\section*{REFERENCES}
\vspace{-15pt}
\bibliography{HIFUUPT}

\providecommand{\noopsort}[1]{}\providecommand{\singleletter}[1]{#1}%
\begin{thebibliography}{33}%
\makeatletter
\providecommand \@ifxundefined [1]{%
 \@ifx{#1\undefined}
}%
\providecommand \@ifnum [1]{%
 \ifnum #1\expandafter \@firstoftwo
 \else \expandafter \@secondoftwo
 \fi
}%
\providecommand \@ifx [1]{%
 \ifx #1\expandafter \@firstoftwo
 \else \expandafter \@secondoftwo
 \fi
}%
\providecommand \natexlab [1]{#1}%
\providecommand \enquote  [1]{``#1''}%
\providecommand \bibnamefont  [1]{#1}%
\providecommand \bibfnamefont [1]{#1}%
\providecommand \citenamefont [1]{#1}%
\providecommand \href@noop [0]{\@secondoftwo}%
\providecommand \href [0]{\begingroup \@sanitize@url \@href}%
\providecommand \@href[1]{\@@startlink{#1}\@@href}%
\providecommand \@@href[1]{\endgroup#1\@@endlink}%
\providecommand \@sanitize@url [0]{\catcode `\\12\catcode `\$12\catcode
  `\&12\catcode `\#12\catcode `\^12\catcode `\_12\catcode `\%12\relax}%
\providecommand \@@startlink[1]{}%
\providecommand \@@endlink[0]{}%
\providecommand \url  [0]{\begingroup\@sanitize@url \@url }%
\providecommand \@url [1]{\endgroup\@href {#1}{\urlprefix }}%
\providecommand \urlprefix  [0]{URL }%
\providecommand \Eprint [0]{\href }%
\providecommand \doibase [0]{http://dx.doi.org/}%
\providecommand \selectlanguage [0]{\@gobble}%
\providecommand \bibinfo  [0]{\@secondoftwo}%
\providecommand \bibfield  [0]{\@secondoftwo}%
\providecommand \translation [1]{[#1]}%
\providecommand \BibitemOpen [0]{}%
\providecommand \bibitemStop [0]{}%
\providecommand \bibitemNoStop [0]{.\EOS\space}%
\providecommand \EOS [0]{\spacefactor3000\relax}%
\providecommand \BibitemShut  [1]{\csname bibitem#1\endcsname}%
\let\auto@bib@innerbib\@empty
\bibitem [{\citenamefont {Cochran}, \citenamefont {Kadaba},\ and\ \citenamefont
  {Palmieri}(1988)}]{cochran1988external}%
  \BibitemOpen
  \bibfield  {author} {\bibinfo {author} {\bibfnamefont {G.~V.}\ \bibnamefont
  {Cochran}}, \bibinfo {author} {\bibfnamefont {M.~P.}\ \bibnamefont {Kadaba}},
  \ and\ \bibinfo {author} {\bibfnamefont {V.~R.}\ \bibnamefont {Palmieri}},\
  }\href@noop {} {\bibfield  {journal} {\bibinfo  {journal} {Journal of
  orthopaedic research}\ }\textbf {\bibinfo {volume} {6}},\ \bibinfo {pages}
  {145} (\bibinfo {year} {1988})}\BibitemShut {NoStop}%
\bibitem [{\citenamefont {Shi}, \citenamefont {Wang},\ and\ \citenamefont
  {Lee}(2016)}]{shi2016mems}%
  \BibitemOpen
  \bibfield  {author} {\bibinfo {author} {\bibfnamefont {Q.}~\bibnamefont
  {Shi}}, \bibinfo {author} {\bibfnamefont {T.}~\bibnamefont {Wang}}, \ and\
  \bibinfo {author} {\bibfnamefont {C.}~\bibnamefont {Lee}},\ }\href@noop {}
  {\bibfield  {journal} {\bibinfo  {journal} {Scientific reports}\ }\textbf
  {\bibinfo {volume} {6}},\ \bibinfo {pages} {24946} (\bibinfo {year}
  {2016})}\BibitemShut {NoStop}%
\bibitem [{\citenamefont {Bao}\ \emph {et~al.}(2008)\citenamefont {Bao},
  \citenamefont {Biederman}, \citenamefont {Sherrit}, \citenamefont {Badescu},
  \citenamefont {Bar-Cohen}, \citenamefont {Jones}, \citenamefont {Aldrich},\
  and\ \citenamefont {Chang}}]{bao2008high}%
  \BibitemOpen
  \bibfield  {author} {\bibinfo {author} {\bibfnamefont {X.}~\bibnamefont
  {Bao}}, \bibinfo {author} {\bibfnamefont {W.}~\bibnamefont {Biederman}},
  \bibinfo {author} {\bibfnamefont {S.}~\bibnamefont {Sherrit}}, \bibinfo
  {author} {\bibfnamefont {M.}~\bibnamefont {Badescu}}, \bibinfo {author}
  {\bibfnamefont {Y.}~\bibnamefont {Bar-Cohen}}, \bibinfo {author}
  {\bibfnamefont {C.}~\bibnamefont {Jones}}, \bibinfo {author} {\bibfnamefont
  {J.}~\bibnamefont {Aldrich}}, \ and\ \bibinfo {author} {\bibfnamefont
  {Z.}~\bibnamefont {Chang}},\ }in\ \href@noop {} {\emph {\bibinfo {booktitle}
  {Industrial and Commercial Applications of Smart Structures Technologies
  2008}}},\ Vol.\ \bibinfo {volume} {6930}\ (\bibinfo {organization}
  {International Society for Optics and Photonics},\ \bibinfo {year} {2008})\
  p.\ \bibinfo {pages} {69300Z}\BibitemShut {NoStop}%
\bibitem [{\citenamefont {Graham}, \citenamefont {Neasham},\ and\ \citenamefont
  {Sharif}(2011)}]{graham2011investigation}%
  \BibitemOpen
  \bibfield  {author} {\bibinfo {author} {\bibfnamefont {D.~J.}\ \bibnamefont
  {Graham}}, \bibinfo {author} {\bibfnamefont {J.~A.}\ \bibnamefont {Neasham}},
  \ and\ \bibinfo {author} {\bibfnamefont {B.~S.}\ \bibnamefont {Sharif}},\
  }\href@noop {} {\bibfield  {journal} {\bibinfo  {journal} {IEEE Transactions
  on industrial electronics}\ }\textbf {\bibinfo {volume} {58}},\ \bibinfo
  {pages} {4972} (\bibinfo {year} {2011})}\BibitemShut {NoStop}%
\bibitem [{\citenamefont {Hu}\ \emph {et~al.}(2003)\citenamefont {Hu},
  \citenamefont {Zhang}, \citenamefont {Yang},\ and\ \citenamefont
  {Jiang}}]{hu2003transmitting}%
  \BibitemOpen
  \bibfield  {author} {\bibinfo {author} {\bibfnamefont {Y.}~\bibnamefont
  {Hu}}, \bibinfo {author} {\bibfnamefont {X.}~\bibnamefont {Zhang}}, \bibinfo
  {author} {\bibfnamefont {J.}~\bibnamefont {Yang}}, \ and\ \bibinfo {author}
  {\bibfnamefont {Q.}~\bibnamefont {Jiang}},\ }\href@noop {} {\bibfield
  {journal} {\bibinfo  {journal} {IEEE Transactions on Ultrasonics,
  Ferroelectrics, and Frequency Control}\ }\textbf {\bibinfo {volume} {50}},\
  \bibinfo {pages} {773} (\bibinfo {year} {2003})}\BibitemShut {NoStop}%
\bibitem [{\citenamefont {Kawanabe}\ \emph {et~al.}(2001)\citenamefont
  {Kawanabe}, \citenamefont {Katane}, \citenamefont {Saotome}, \citenamefont
  {Saito},\ and\ \citenamefont {Kobayashi}}]{kawanabe2001power}%
  \BibitemOpen
  \bibfield  {author} {\bibinfo {author} {\bibfnamefont {H.}~\bibnamefont
  {Kawanabe}}, \bibinfo {author} {\bibfnamefont {T.}~\bibnamefont {Katane}},
  \bibinfo {author} {\bibfnamefont {H.}~\bibnamefont {Saotome}}, \bibinfo
  {author} {\bibfnamefont {O.}~\bibnamefont {Saito}}, \ and\ \bibinfo {author}
  {\bibfnamefont {K.}~\bibnamefont {Kobayashi}},\ }\href@noop {} {\bibfield
  {journal} {\bibinfo  {journal} {Japanese journal of applied physics}\
  }\textbf {\bibinfo {volume} {40}},\ \bibinfo {pages} {3865} (\bibinfo {year}
  {2001})}\BibitemShut {NoStop}%
\bibitem [{\citenamefont {Sanni}, \citenamefont {Vilches},\ and\ \citenamefont
  {Toumazou}(2012)}]{sanni2012inductive}%
  \BibitemOpen
  \bibfield  {author} {\bibinfo {author} {\bibfnamefont {A.}~\bibnamefont
  {Sanni}}, \bibinfo {author} {\bibfnamefont {A.}~\bibnamefont {Vilches}}, \
  and\ \bibinfo {author} {\bibfnamefont {C.}~\bibnamefont {Toumazou}},\
  }\href@noop {} {\bibfield  {journal} {\bibinfo  {journal} {IEEE transactions
  on biomedical circuits and systems}\ }\textbf {\bibinfo {volume} {6}},\
  \bibinfo {pages} {297} (\bibinfo {year} {2012})}\BibitemShut {NoStop}%
\bibitem [{\citenamefont {Shahab}\ and\ \citenamefont
  {Erturk}(2014)}]{shahab2014contactless}%
  \BibitemOpen
  \bibfield  {author} {\bibinfo {author} {\bibfnamefont {S.}~\bibnamefont
  {Shahab}}\ and\ \bibinfo {author} {\bibfnamefont {A.}~\bibnamefont
  {Erturk}},\ }\href@noop {} {\bibfield  {journal} {\bibinfo  {journal} {Smart
  Materials and Structures}\ }\textbf {\bibinfo {volume} {23}},\ \bibinfo
  {pages} {125032} (\bibinfo {year} {2014})}\BibitemShut {NoStop}%
\bibitem [{\citenamefont {Ozeri}\ \emph {et~al.}(2010)\citenamefont {Ozeri},
  \citenamefont {Shmilovitz}, \citenamefont {Singer},\ and\ \citenamefont
  {Wang}}]{ozeri2010ultrasonic}%
  \BibitemOpen
  \bibfield  {author} {\bibinfo {author} {\bibfnamefont {S.}~\bibnamefont
  {Ozeri}}, \bibinfo {author} {\bibfnamefont {D.}~\bibnamefont {Shmilovitz}},
  \bibinfo {author} {\bibfnamefont {S.}~\bibnamefont {Singer}}, \ and\ \bibinfo
  {author} {\bibfnamefont {C.-C.}\ \bibnamefont {Wang}},\ }\href@noop {}
  {\bibfield  {journal} {\bibinfo  {journal} {Ultrasonics}\ }\textbf {\bibinfo
  {volume} {50}},\ \bibinfo {pages} {666} (\bibinfo {year} {2010})}\BibitemShut
  {NoStop}%
\bibitem [{\citenamefont {Shahab}, \citenamefont {Gray},\ and\ \citenamefont
  {Erturk}(2015)}]{shahab2015ultrasonic}%
  \BibitemOpen
  \bibfield  {author} {\bibinfo {author} {\bibfnamefont {S.}~\bibnamefont
  {Shahab}}, \bibinfo {author} {\bibfnamefont {M.}~\bibnamefont {Gray}}, \ and\
  \bibinfo {author} {\bibfnamefont {A.}~\bibnamefont {Erturk}},\ }\href@noop {}
  {\bibfield  {journal} {\bibinfo  {journal} {Journal of Applied Physics}\
  }\textbf {\bibinfo {volume} {117}},\ \bibinfo {pages} {104903} (\bibinfo
  {year} {2015})}\BibitemShut {NoStop}%
\bibitem [{\citenamefont {Meesala}, \citenamefont {Hajj},\ and\ \citenamefont
  {Shahab}(2020)}]{meesala2020modeling}%
  \BibitemOpen
  \bibfield  {author} {\bibinfo {author} {\bibfnamefont {V.~C.}\ \bibnamefont
  {Meesala}}, \bibinfo {author} {\bibfnamefont {M.~R.}\ \bibnamefont {Hajj}}, \
  and\ \bibinfo {author} {\bibfnamefont {S.}~\bibnamefont {Shahab}},\
  }\href@noop {} {\bibfield  {journal} {\bibinfo  {journal} {Nonlinear
  Dynamics}\ }\textbf {\bibinfo {volume} {99}},\ \bibinfo {pages} {249}
  (\bibinfo {year} {2020})}\BibitemShut {NoStop}%
\bibitem [{\citenamefont {Meesala}\ \emph {et~al.}(2020)\citenamefont
  {Meesala}, \citenamefont {Ragab}, \citenamefont {Hajj},\ and\ \citenamefont
  {Shahab}}]{meesala2020acoustic}%
  \BibitemOpen
  \bibfield  {author} {\bibinfo {author} {\bibfnamefont {V.~C.}\ \bibnamefont
  {Meesala}}, \bibinfo {author} {\bibfnamefont {S.}~\bibnamefont {Ragab}},
  \bibinfo {author} {\bibfnamefont {M.~R.}\ \bibnamefont {Hajj}}, \ and\
  \bibinfo {author} {\bibfnamefont {S.}~\bibnamefont {Shahab}},\ }\href@noop {}
  {\bibfield  {journal} {\bibinfo  {journal} {Journal of Sound and Vibration}\
  ,\ \bibinfo {pages} {115255}} (\bibinfo {year} {2020})}\BibitemShut {NoStop}%
\bibitem [{\citenamefont {Roes}\ \emph {et~al.}(2012)\citenamefont {Roes},
  \citenamefont {Duarte}, \citenamefont {Hendrix},\ and\ \citenamefont
  {Lomonova}}]{roes2012acoustic}%
  \BibitemOpen
  \bibfield  {author} {\bibinfo {author} {\bibfnamefont {M.~G.}\ \bibnamefont
  {Roes}}, \bibinfo {author} {\bibfnamefont {J.~L.}\ \bibnamefont {Duarte}},
  \bibinfo {author} {\bibfnamefont {M.~A.}\ \bibnamefont {Hendrix}}, \ and\
  \bibinfo {author} {\bibfnamefont {E.~A.}\ \bibnamefont {Lomonova}},\
  }\href@noop {} {\bibfield  {journal} {\bibinfo  {journal} {IEEE Transactions
  on Industrial Electronics}\ }\textbf {\bibinfo {volume} {60}},\ \bibinfo
  {pages} {242} (\bibinfo {year} {2012})}\BibitemShut {NoStop}%
\bibitem [{\citenamefont {Awal}\ \emph {et~al.}(2016)\citenamefont {Awal},
  \citenamefont {Jusoh}, \citenamefont {Sabapathy}, \citenamefont {Kamarudin},\
  and\ \citenamefont {Rahim}}]{awal2016state}%
  \BibitemOpen
  \bibfield  {author} {\bibinfo {author} {\bibfnamefont {M.~R.}\ \bibnamefont
  {Awal}}, \bibinfo {author} {\bibfnamefont {M.}~\bibnamefont {Jusoh}},
  \bibinfo {author} {\bibfnamefont {T.}~\bibnamefont {Sabapathy}}, \bibinfo
  {author} {\bibfnamefont {M.~R.}\ \bibnamefont {Kamarudin}}, \ and\ \bibinfo
  {author} {\bibfnamefont {R.~A.}\ \bibnamefont {Rahim}},\ }\href@noop {}
  {\bibfield  {journal} {\bibinfo  {journal} {International Journal of Antennas
  and Propagation}\ }\textbf {\bibinfo {volume} {2016}} (\bibinfo {year}
  {2016})}\BibitemShut {NoStop}%
\bibitem [{\citenamefont {Kim}\ \emph {et~al.}(2012)\citenamefont {Kim},
  \citenamefont {Ho}, \citenamefont {Chen},\ and\ \citenamefont
  {Poon}}]{kim2012wireless}%
  \BibitemOpen
  \bibfield  {author} {\bibinfo {author} {\bibfnamefont {S.}~\bibnamefont
  {Kim}}, \bibinfo {author} {\bibfnamefont {J.~S.}\ \bibnamefont {Ho}},
  \bibinfo {author} {\bibfnamefont {L.~Y.}\ \bibnamefont {Chen}}, \ and\
  \bibinfo {author} {\bibfnamefont {A.~S.}\ \bibnamefont {Poon}},\ }\href@noop
  {} {\bibfield  {journal} {\bibinfo  {journal} {Applied Physics Letters}\
  }\textbf {\bibinfo {volume} {101}},\ \bibinfo {pages} {073701} (\bibinfo
  {year} {2012})}\BibitemShut {NoStop}%
\bibitem [{\citenamefont {Li}\ \emph {et~al.}(2012)\citenamefont {Li},
  \citenamefont {Yuan}, \citenamefont {Yang}, \citenamefont {Liu},\ and\
  \citenamefont {Zhang}}]{li2012magnetic}%
  \BibitemOpen
  \bibfield  {author} {\bibinfo {author} {\bibfnamefont {X.}~\bibnamefont
  {Li}}, \bibinfo {author} {\bibfnamefont {Q.}~\bibnamefont {Yuan}}, \bibinfo
  {author} {\bibfnamefont {T.}~\bibnamefont {Yang}}, \bibinfo {author}
  {\bibfnamefont {J.}~\bibnamefont {Liu}}, \ and\ \bibinfo {author}
  {\bibfnamefont {H.}~\bibnamefont {Zhang}},\ }\href@noop {} {\bibfield
  {journal} {\bibinfo  {journal} {Journal of Applied Physics}\ }\textbf
  {\bibinfo {volume} {111}},\ \bibinfo {pages} {07E734} (\bibinfo {year}
  {2012})}\BibitemShut {NoStop}%
\bibitem [{\citenamefont {Kar}\ \emph {et~al.}(2015)\citenamefont {Kar},
  \citenamefont {Nayak}, \citenamefont {Bhuyan},\ and\ \citenamefont
  {Mishra}}]{kar2015bi}%
  \BibitemOpen
  \bibfield  {author} {\bibinfo {author} {\bibfnamefont {D.~P.}\ \bibnamefont
  {Kar}}, \bibinfo {author} {\bibfnamefont {P.~P.}\ \bibnamefont {Nayak}},
  \bibinfo {author} {\bibfnamefont {S.}~\bibnamefont {Bhuyan}}, \ and\ \bibinfo
  {author} {\bibfnamefont {D.}~\bibnamefont {Mishra}},\ }\href@noop {}
  {\bibfield  {journal} {\bibinfo  {journal} {Applied Physics Letters}\
  }\textbf {\bibinfo {volume} {107}},\ \bibinfo {pages} {133901} (\bibinfo
  {year} {2015})}\BibitemShut {NoStop}%
\bibitem [{\citenamefont {Cui}\ and\ \citenamefont
  {Yang}(2011)}]{cui2011enhanced}%
  \BibitemOpen
  \bibfield  {author} {\bibinfo {author} {\bibfnamefont {H.}~\bibnamefont
  {Cui}}\ and\ \bibinfo {author} {\bibfnamefont {X.}~\bibnamefont {Yang}},\
  }\href@noop {} {\bibfield  {journal} {\bibinfo  {journal} {Applied physics
  letters}\ }\textbf {\bibinfo {volume} {99}},\ \bibinfo {pages} {231113}
  (\bibinfo {year} {2011})}\BibitemShut {NoStop}%
\bibitem [{\citenamefont {Cui}, \citenamefont {Zhang},\ and\ \citenamefont
  {Yang}(2013)}]{cui2013laser}%
  \BibitemOpen
  \bibfield  {author} {\bibinfo {author} {\bibfnamefont {H.}~\bibnamefont
  {Cui}}, \bibinfo {author} {\bibfnamefont {T.}~\bibnamefont {Zhang}}, \ and\
  \bibinfo {author} {\bibfnamefont {X.}~\bibnamefont {Yang}},\ }\href@noop {}
  {\bibfield  {journal} {\bibinfo  {journal} {Applied physics letters}\
  }\textbf {\bibinfo {volume} {102}},\ \bibinfo {pages} {133702} (\bibinfo
  {year} {2013})}\BibitemShut {NoStop}%
\bibitem [{\citenamefont {Bhargava}\ \emph {et~al.}(2017)\citenamefont
  {Bhargava}, \citenamefont {Peng}, \citenamefont {Stieg}, \citenamefont
  {Mirzaeifar},\ and\ \citenamefont {Shahab}}]{bhargava2017focused}%
  \BibitemOpen
  \bibfield  {author} {\bibinfo {author} {\bibfnamefont {A.}~\bibnamefont
  {Bhargava}}, \bibinfo {author} {\bibfnamefont {K.}~\bibnamefont {Peng}},
  \bibinfo {author} {\bibfnamefont {J.}~\bibnamefont {Stieg}}, \bibinfo
  {author} {\bibfnamefont {R.}~\bibnamefont {Mirzaeifar}}, \ and\ \bibinfo
  {author} {\bibfnamefont {S.}~\bibnamefont {Shahab}},\ }\href@noop {}
  {\bibfield  {journal} {\bibinfo  {journal} {RSC advances}\ }\textbf {\bibinfo
  {volume} {7}},\ \bibinfo {pages} {45452} (\bibinfo {year}
  {2017})}\BibitemShut {NoStop}%
\bibitem [{\citenamefont {Rudenko}, \citenamefont {Hedberg},\ and\
  \citenamefont {Enflo}(2001)}]{rudenko2001nonlinear}%
  \BibitemOpen
  \bibfield  {author} {\bibinfo {author} {\bibfnamefont {O.}~\bibnamefont
  {Rudenko}}, \bibinfo {author} {\bibfnamefont {C.}~\bibnamefont {Hedberg}}, \
  and\ \bibinfo {author} {\bibfnamefont {B.}~\bibnamefont {Enflo}},\
  }\href@noop {} {\bibfield  {journal} {\bibinfo  {journal} {Acoustical
  Physics}\ }\textbf {\bibinfo {volume} {47}},\ \bibinfo {pages} {452}
  (\bibinfo {year} {2001})}\BibitemShut {NoStop}%
\bibitem [{\citenamefont {Vanhille}\ and\ \citenamefont
  {Campos-Pozuelo}(2005)}]{vanhille2005numerical}%
  \BibitemOpen
  \bibfield  {author} {\bibinfo {author} {\bibfnamefont {C.}~\bibnamefont
  {Vanhille}}\ and\ \bibinfo {author} {\bibfnamefont {C.}~\bibnamefont
  {Campos-Pozuelo}},\ }\href@noop {} {\bibfield  {journal} {\bibinfo  {journal}
  {Ultrasonics}\ }\textbf {\bibinfo {volume} {43}},\ \bibinfo {pages} {652}
  (\bibinfo {year} {2005})}\BibitemShut {NoStop}%
\bibitem [{\citenamefont {Coppens}\ and\ \citenamefont
  {Sanders}(1968)}]{coppens1968finite}%
  \BibitemOpen
  \bibfield  {author} {\bibinfo {author} {\bibfnamefont {A.~B.}\ \bibnamefont
  {Coppens}}\ and\ \bibinfo {author} {\bibfnamefont {J.~V.}\ \bibnamefont
  {Sanders}},\ }\href@noop {} {\bibfield  {journal} {\bibinfo  {journal} {The
  Journal of the Acoustical Society of America}\ }\textbf {\bibinfo {volume}
  {43}},\ \bibinfo {pages} {516} (\bibinfo {year} {1968})}\BibitemShut
  {NoStop}%
\bibitem [{\citenamefont {Andrade}, \citenamefont {Bernassau},\ and\
  \citenamefont {Adamowski}(2016)}]{andrade2016acoustic}%
  \BibitemOpen
  \bibfield  {author} {\bibinfo {author} {\bibfnamefont {M.~A.}\ \bibnamefont
  {Andrade}}, \bibinfo {author} {\bibfnamefont {A.~L.}\ \bibnamefont
  {Bernassau}}, \ and\ \bibinfo {author} {\bibfnamefont {J.~C.}\ \bibnamefont
  {Adamowski}},\ }\href@noop {} {\bibfield  {journal} {\bibinfo  {journal}
  {Applied Physics Letters}\ }\textbf {\bibinfo {volume} {109}},\ \bibinfo
  {pages} {044101} (\bibinfo {year} {2016})}\BibitemShut {NoStop}%
\bibitem [{\citenamefont {Bassindale}\ \emph {et~al.}(2014)\citenamefont
  {Bassindale}, \citenamefont {Phillips}, \citenamefont {Barnes},\ and\
  \citenamefont {Drinkwater}}]{bassindale2014measurements}%
  \BibitemOpen
  \bibfield  {author} {\bibinfo {author} {\bibfnamefont {P.}~\bibnamefont
  {Bassindale}}, \bibinfo {author} {\bibfnamefont {D.}~\bibnamefont
  {Phillips}}, \bibinfo {author} {\bibfnamefont {A.}~\bibnamefont {Barnes}}, \
  and\ \bibinfo {author} {\bibfnamefont {B.}~\bibnamefont {Drinkwater}},\
  }\href@noop {} {\bibfield  {journal} {\bibinfo  {journal} {Applied Physics
  Letters}\ }\textbf {\bibinfo {volume} {104}},\ \bibinfo {pages} {163504}
  (\bibinfo {year} {2014})}\BibitemShut {NoStop}%
\bibitem [{\citenamefont {Hamilton}, \citenamefont {Blackstock}\ \emph
  {et~al.}(1998)\citenamefont {Hamilton}, \citenamefont {Blackstock} \emph
  {et~al.}}]{hamilton1998nonlinear}%
  \BibitemOpen
  \bibfield  {author} {\bibinfo {author} {\bibfnamefont {M.~F.}\ \bibnamefont
  {Hamilton}}, \bibinfo {author} {\bibfnamefont {D.~T.}\ \bibnamefont
  {Blackstock}},  \emph {et~al.},\ }\href@noop {} {\emph {\bibinfo {title}
  {Nonlinear acoustics}}},\ Vol.\ \bibinfo {volume} {237}\ (\bibinfo
  {publisher} {Academic press San Diego},\ \bibinfo {year} {1998})\BibitemShut
  {NoStop}%
\bibitem [{\citenamefont {Hart}\ and\ \citenamefont
  {Hamilton}(1988)}]{hart1988nonlinear}%
  \BibitemOpen
  \bibfield  {author} {\bibinfo {author} {\bibfnamefont {T.~S.}\ \bibnamefont
  {Hart}}\ and\ \bibinfo {author} {\bibfnamefont {M.~F.}\ \bibnamefont
  {Hamilton}},\ }\href@noop {} {\bibfield  {journal} {\bibinfo  {journal} {The
  Journal of the Acoustical Society of America}\ }\textbf {\bibinfo {volume}
  {84}},\ \bibinfo {pages} {1488} (\bibinfo {year} {1988})}\BibitemShut
  {NoStop}%
\bibitem [{\citenamefont {Canney}\ \emph {et~al.}(2008)\citenamefont {Canney},
  \citenamefont {Bailey}, \citenamefont {Crum}, \citenamefont {Khokhlova},\
  and\ \citenamefont {Sapozhnikov}}]{canney2008acoustic}%
  \BibitemOpen
  \bibfield  {author} {\bibinfo {author} {\bibfnamefont {M.~S.}\ \bibnamefont
  {Canney}}, \bibinfo {author} {\bibfnamefont {M.~R.}\ \bibnamefont {Bailey}},
  \bibinfo {author} {\bibfnamefont {L.~A.}\ \bibnamefont {Crum}}, \bibinfo
  {author} {\bibfnamefont {V.~A.}\ \bibnamefont {Khokhlova}}, \ and\ \bibinfo
  {author} {\bibfnamefont {O.~A.}\ \bibnamefont {Sapozhnikov}},\ }\href@noop {}
  {\bibfield  {journal} {\bibinfo  {journal} {The Journal of the Acoustical
  Society of America}\ }\textbf {\bibinfo {volume} {124}},\ \bibinfo {pages}
  {2406} (\bibinfo {year} {2008})}\BibitemShut {NoStop}%
\bibitem [{\citenamefont {Duck}(2002)}]{duck2002nonlinear}%
  \BibitemOpen
  \bibfield  {author} {\bibinfo {author} {\bibfnamefont {F.~A.}\ \bibnamefont
  {Duck}},\ }\href@noop {} {\bibfield  {journal} {\bibinfo  {journal}
  {Ultrasound in medicine \& biology}\ }\textbf {\bibinfo {volume} {28}},\
  \bibinfo {pages} {1} (\bibinfo {year} {2002})}\BibitemShut {NoStop}%
\bibitem [{\citenamefont {Muir}\ and\ \citenamefont
  {Carstensen}(1980)}]{muir1980prediction}%
  \BibitemOpen
  \bibfield  {author} {\bibinfo {author} {\bibfnamefont {T.}~\bibnamefont
  {Muir}}\ and\ \bibinfo {author} {\bibfnamefont {E.}~\bibnamefont
  {Carstensen}},\ }\href@noop {} {\bibfield  {journal} {\bibinfo  {journal}
  {Ultrasound in medicine \& biology}\ }\textbf {\bibinfo {volume} {6}},\
  \bibinfo {pages} {345} (\bibinfo {year} {1980})}\BibitemShut {NoStop}%
\bibitem [{\citenamefont {Fahy}\ and\ \citenamefont
  {Gardonio}(2007)}]{fahy2007sound}%
  \BibitemOpen
  \bibfield  {author} {\bibinfo {author} {\bibfnamefont {F.~J.}\ \bibnamefont
  {Fahy}}\ and\ \bibinfo {author} {\bibfnamefont {P.}~\bibnamefont
  {Gardonio}},\ }\href@noop {} {\emph {\bibinfo {title} {Sound and structural
  vibration: radiation, transmission and response}}}\ (\bibinfo  {publisher}
  {Elsevier},\ \bibinfo {year} {2007})\BibitemShut {NoStop}%
\bibitem [{\citenamefont {Leo}(2007)}]{leo2007engineering}%
  \BibitemOpen
  \bibfield  {author} {\bibinfo {author} {\bibfnamefont {D.~J.}\ \bibnamefont
  {Leo}},\ }\href@noop {} {\emph {\bibinfo {title} {Engineering analysis of
  smart material systems}}}\ (\bibinfo  {publisher} {John Wiley \& Sons},\
  \bibinfo {year} {2007})\BibitemShut {NoStop}%
\bibitem [{\citenamefont {Bhargava}\ and\ \citenamefont
  {Shahab}(2020)}]{bhargava2020contactless}%
  \BibitemOpen
  \bibfield  {author} {\bibinfo {author} {\bibfnamefont {A.}~\bibnamefont
  {Bhargava}}\ and\ \bibinfo {author} {\bibfnamefont {S.}~\bibnamefont
  {Shahab}},\ }\href@noop {} {\bibfield  {journal} {\bibinfo  {journal} {arXiv
  preprint arXiv:2006.08054}\ } (\bibinfo {year} {2020})}\BibitemShut {NoStop}%
\end{thebibliography}%

\end{document}